# Snapshot Multispectral Imaging Using a Diffractive Optical Network

Deniz Mengu[1,2,3], Anika Tabassum[1,2,3], Mona Jarrahi[1,2,3], Aydogan Ozcan[1,2,3,*]

[1] Electrical and Computer Engineering Department, University of California, Los Angeles, CA, 90095, USA

[2] Bioengineering Department, University of California, Los Angeles, CA, 90095, USA

[3] California NanoSystems Institute, University of California, Los Angeles, CA, 90095, USA

[§] Equal contribution

* Corresponding author: ozcan@ucla.edu

## Abstract

Multispectral imaging has been used for numerous applications in e.g., environmental monitoring, aerospace, defense, and biomedicine. Here, we present a diffractive optical network-based multispectral imaging system trained using deep learning to create a virtual spectral filter array at the output image field-of-view. This diffractive multispectral imager performs spatially-coherent imaging over a large spectrum, and at the same time, routes a pre-determined set of spectral channels onto an array of pixels at the output plane, converting a monochrome focal plane array or image sensor into a multispectral imaging device without any spectral filters or image recovery algorithms. Furthermore, the spectral responsivity of this diffractive multispectral imager is not sensitive to input polarization states. Through numerical simulations, we present different diffractive network designs that achieve snapshot multispectral imaging with 4, 9 and 16 unique spectral bands within the visible spectrum, based on passive spatially-structured diffractive surfaces, with a compact design that axially spans ~$72\lambda_m$, where $\lambda_m$ is the mean wavelength of the spectral band of interest. Moreover, we experimentally demonstrate a diffractive multispectral imager based on a 3D-printed diffractive network that creates at its output image plane a spatially-repeating virtual spectral filter array with 2x2=4 unique bands at terahertz spectrum. Due to their compact form factor and computation-free, power-efficient and polarization-insensitive forward operation, diffractive multispectral imagers can be transformative for various imaging and sensing applications and be used at different parts of the electromagnetic spectrum where high-density and wide-area multispectral pixel arrays are not widely available.

## 1. Introduction

Multispectral imaging has been an instrumental tool for major advances in various fields, including environmental monitoring[1], astronomy[2–4], agricultural sciences[5,6], biological imaging[7–9], medical diagnostics[10,11], and food quality control[12,13] among many others[14–20]. One of the simplest ways to achieve multispectral imaging is to sacrifice the image acquisition time in favor of the spectral information by capturing multiple shots of a scene while changing the spectral filter in front of a monochrome camera[21]. Another traditional form of multispectral imaging relies on push-broom scanning of a one-dimensional detector array across the field-of-view (FOV)[22]. While these multispectral imaging techniques provide sufficient spectral and spatial resolution, they suffer from relatively long data acquisition times, hindering their use in real-time imaging applications[23]. An alternative solution that allows simultaneous collection of the spatial and spectral information is to split the optical waves emanating from the input FOV onto different optical paths each containing a different spectral filter, followed by a 2D monochrome image sensor array[24,25]. However, this approach often leads to more complex and bulky optical systems since it requires the use of multiple focal-plane arrays, one for each band, along with other optical components.

Modern-day snapshot spectral imaging systems often use coded apertures in conjunction with computational image recovery algorithms to digitally mitigate these shortcomings of traditional multispectral imaging systems. One of the earliest forms of coded aperture snapshot spectral imaging used a binary spatial aperture function imaged onto a dispersive optical element through relay optics, encoding both the spatial and spectral features contained within the input FOV into an intensity pattern collected by a monochrome focal-plane array[26]. Since this initial proof-of-concept demonstration, various improvements have been reported on coded aperture-based snapshot spectral imaging systems based on, e.g., the use of color-coded apertures[27], compressive sensing techniques[28–31] and others[32]. On the other hand, these systems still require the use of optical relay systems and dispersive optical elements such as prisms, and diffractive elements, resulting in relatively bulky form factors; furthermore, their frame rate is often limited by the computationally intense iterative recovery algorithms that are used to digitally retrieve the multispectral image cube from the raw data. Recent studies have also reported using diffractive lens designs, addressing the form factor limitations of multispectral imaging systems[33–35]. These approaches provide restricted spatial and spectral encoding capabilities due to their limited degrees of freedom without coded apertures, causing relatively poor spectral resolution. Recent work also demonstrated the use of feedforward deep neural networks to achieve better image reconstruction quality, addressing some of the limitations imposed by the iterative reconstruction algorithms typically employed in multispectral imaging and sensing[36–38]. On the other hand, deep learning-enabled computational multispectral imagers require access to powerful graphics processing units (GPUs)[39] for rapid inference of each spectral image cube and rely on training data acquisition or a calibration process to characterize their point spread functions[33].

With the development of high-resolution image sensor-arrays, it has become more practical to compromise spatial resolution to collect richer spectral information. The most ubiquitous form of a relatively primitive spectral imaging device designed around this trade-off is a color camera based on the Bayer filters (R, G, B channels, representing the red, green and blue spectral bands, respectively). The traditional RGB color image sensor is based on a periodically repeating array of 2×2 pixels, with each subpixel containing an absorptive spectral filter (also known as the Bayer filters) that transmits the red, green, or blue wavelengths while partially blocking the others. Despite its frequent use in various imaging applications, there has been a tremendous effort to develop better alternatives to these absorptive filters that suffer from a relatively high-cross talk, low power efficiency, and poor color representation[40]. Towards this end, numerous engineered optical material structures have been explored, including plasmonic antennas[41], dielectric metasurfaces[42–46] and 3D porous materials[47–49]. While the intrinsic losses associated with metallic nanostructures limit their optical efficiency, multispectral imager designs based on dielectric metasurfaces and 3D porous compound optical elements have been reported to achieve higher power efficiencies with lower color crosstalk[40]. However, these structured material-based approaches, including various metamaterial designs, were all limited to 4 or fewer spectral channels, and did not demonstrate a large array of spectral filters for multispectral imaging. Independent from these spectral filtering techniques based on optimized meta-designs, increasing the number of unique spectral channels in conventional multispectral filters was also demonstrated, which, in general, poses various design and implementation challenges for scale-up[23,50].

Here, we introduce the design of a snapshot multispectral imager that is based on a diffractive optical network (also known as $D^2NN$, diffractive deep neural network[51–60]) and demonstrate its performance with 4 (2×2), 9 (3×3) and 16 (4×4) unique spectral bands that are periodically repeating at the output image FOV to form a *virtual* multispectral filter array. This diffractive network-based multispectral imager (Fig. 1) is trained to project the spatial information of an object onto a grid of virtual pixels, with each one carrying the information of a pre-determined spectral band, performing snapshot multispectral imaging via engineered diffraction of light through passive transmissive layers that axially span ~$72\lambda_m$, where $\lambda_m$ is the mean wavelength of the entire spectral band of interest. This unique multispectral imager design based on diffractive optical networks achieves two tasks simultaneously: (1) its acts as a broadband spatially-coherent relay optics achieving the optical imaging task between the input and the output FOVs over a wide spectral range; and (2) it spatially separates the input spectral channels into distinct pixels at the same output image plane, serving as a *virtual spectral filter array* that preserves the spatial information of the scene/object, instantaneously yielding an image cube without image reconstruction algorithms, except the standard demosaicing of the virtual filter array pixels. Stated differently, we demonstrate diffractive optical networks that virtually convert a monochrome focal plane array or an image sensor into a snapshot multispectral imaging device without the need for conventional spectral filters.

We present different numerical diffractive network designs that achieve multispectral coherent imaging with 4, 9 and 16 unique spectral bands within the visible spectrum based on passive diffractive layers that are laterally engineered at a feature size of ~225 nm, spanning ~43 µm in the axial direction from the first layer to the last, forming a compact and scalable design. Our numerical analyses on the spectral signal contrast provided by these diffractive multispectral imagers reveal that for a given array of virtual filter pixels (covering, e.g., 4, 9 and 16 spectral bands), the mean optical power of each one of the targeted spectral bands is approximately an order of magnitude larger compared to the average optical power of the other wavelengths.

Furthermore, we experimentally demonstrate the success of our diffractive multispectral imaging framework using a 3D-printed diffractive network operating at terahertz wavelengths. Targeting peak frequencies at 0.375, 0.400, 0.425 and 0.450 THz, the fabricated diffractive network with 3 structured transmissive layers can successfully route each spectral component onto a corresponding array of virtual pixels at the output image plane, forming a multispectral coherent imager with 4 spectral channels. Although we focused on spatially-coherent multispectral imaging in this work, phase-only diffractive layers can also be optimized using deep learning to create spatially incoherent snapshot multispectral imagers, following the same design principles outlined here. With its compact form factor and snapshot operation without any image cube reconstruction algorithms, we believe the presented diffractive multispectral imaging framework can be transformative in various imaging and sensing applications. Since the presented diffractive multispectral imagers utilize isotropic dielectric materials, their virtual spectral filter arrays are not sensitive to the input polarization state of the illumination light, which provides an additional advantage. Finally, due to its scalability, it can drive the development of multispectral imagers at any part of the electromagnetic spectrum, which would be especially important for bands where high-density and large-format spectral filter arrays are not widely available or too costly.

## 2. Results and Discussion

Figure 1 depicts the optical layout and the forward model of a 5-layer diffractive multispectral imager that can spatially separate $N_B$ distinct spectral bands into a virtual spectral filter array on a monochrome image sensor located at the output image plane; in this illustration of Fig. 1, $N_B = 9$ is shown as an example, although it can be further increased, as will be reported below. The input FOV in Fig. 1 exemplifies a hypothetical object where the amplitude channel of the object's light transmission is composed of intersecting lines, and each line strictly transmits only one wavelength. Such a forward optical transformation can be defined using a linear spatial transformation ($y = x$) between the input and output intensities for each targeted spectral band since image sensors are only sensitive to the incident optical intensity. This indicates that for a diffractive network-based spatially-coherent multispectral imager, there is a phase degree of freedom at the output image plane, making it easier to learn the desired multispectral imaging task through, e.g., deep learning. For a diffractive multispectral imager, as shown in Fig. 1, $N_i$ and $N_o$ indicate the number of effective pixels at the input and output FOVs, respectively, which are dictated by the extent of the input and output FOVs along with the desired spatial resolution (within the diffraction limit)[55,56]. The number of spectral channels ($N_B$) as part of the targeted multispectral imaging design depends on the cross-talk among different spectral bands of the virtual filter array created at the diffractive network output, which will be quantified in our analysis reported below. Although not demonstrated here, in alternative implementations, the diffractive multispectral network can also be placed right behind the image plane of a camera, transferring the multispectral image of an object onto the plane of the monochrome focal plane array, converting an existing monochrome imaging system into a multispectral imager.

To train (and design) our diffractive multispectral imager, we created input objects, where the transmission field amplitude of a given object at each spectral band was represented by an image randomly selected from the 101.6K training images of the EMNIST dataset (see the Methods section). The phase profiles of the five diffractive layers (containing ~0.76 million trainable diffractive features in total) were optimized through the error-backpropagation and stochastic gradient descent using a loss function based on the spatial mean-squared error (MSE) that includes all the desired spectral channels; see the Supplementary Methods section. This deep learning-based optimization used 100 epochs, where the ground truth multispectral output images were generated using the EMNIST dataset randomly assigned to different spectral bands of interest. Figure 2a illustrates the resulting material thickness profiles of a $K = 5$ layer diffractive multispectral imager trained to operate within the visible spectrum, evenly covering the wavelength range from $\lambda_9 = 450$ nm to $\lambda_1 = 700$ nm based on the optical layout shown in Fig. 1, i.e., $\lambda_9 < \lambda_8 < \cdots < \lambda_1$. For simplicity and without loss of generality, we assume the input light spectrum to lie between 450 nm and 700 nm; modern CMOS image sensors cover a slightly wider bandwidth than considered here. The forward optical training model of this diffractive network assumes a monochrome image sensor at the output plane with a pixel size of $0.9\mu m \times 0.9\mu m$ (~$1.28\lambda_1 \times 1.28\lambda_1$), which is typical for today's CMOS image sensor technology widely deployed in, e.g., smartphone cameras[61]. This diffractive design spatially extends ~43 μm in the axial direction (from the first diffractive layer to the last layer), and is optimized to route $N_B = 9$ distinct spectral lines (i.e., 700 nm, 668.75 nm, 637.5 nm, 606.25 nm, 575 nm, 543.75 nm, 512.5 nm, 481.25 nm, and 450 nm) onto a 3×3 monochrome sensor pixel-array, that is repeating in space for snapshot multispectral imaging without any digital image reconstruction algorithm. Without loss of generality, we assumed unit magnification between the object/input FOV and the monochrome image sensor plane (output FOV); hence, the size of the smallest feature size of the input images was set to be $3 \times 0.9\mu m$, i.e., equal to the width of a virtual spectral filter array (3x3).

Following the deep learning-based training and design phase (see the Supplementary Methods section for further details), a multicolor image test set with a total of 2080 distinct objects (never seen during the training) was used to quantify the multispectral imaging performance of the trained diffractive network design. For each object in our blind test set, the field amplitude of the object transmission function at each spectral band was modeled based on an image randomly selected from the test dataset. An example of the imaging results corresponding to a multispectral test object never used during the training is shown in Fig. 2b. Based on the checkerboard-like output intensity patterns synthesized by the diffractive multispectral imager in response to the 2080 different test objects, the spectral image contrast of the diffractive network output can be quantified as shown in Fig. 2c; each row of the matrix in Fig. 2c corresponds to a different illumination wavelength and all the rows sum up to 100% (optical power). Hence, the rows of this matrix represent the percentage of the output optical power that resides within the designated group of virtual pixels for a given wavelength channel, calculated as an average of all the 2080 blind test objects. The columns of the matrix in Fig. 2c, on the other hand, illustrate the signal contrast and the spectral leakage over a given array of virtual spectral filters assigned to a spectral band. Our analyses show that for a given set of virtual spectral pixels assigned to a particular spectral band (a column of the matrix in Fig. 2c), the power of the desired signal band is on average (8.57±1.59)-fold larger compared to the mean power of the other spectral bands (leakage) collected by the same array of virtual spectral filter pixels.

Based on the data shown in Fig. 2c, we see that the performance of the diffractive multispectral imager is inversely proportional to the wavelength. In other words, the diffractive optical network designed using deep learning can route smaller wavelengths onto their corresponding virtual spectral filter locations better than larger wavelengths. A similar conclusion can also be observed in the spectral responsibility curves of the 3x3 virtual filter array, periodically assigned to $N_B = 9$ (see Fig. 3); the responsivity curves of these virtual filter arrays get narrower as the wavelength gets smaller, with the narrowest filter response achieved for $\lambda_9 = 450$ nm. These observations can be explained based on the degrees of freedom available at each wavelength: due to the diffraction limit of light, the *effective* number of trainable diffractive features seen/controlled by larger wavelengths is smaller than the total number of trainable features within the entire diffractive network, $N = 5 \times 392 \times 392$. For example, a given diffractive layer depicted in Fig. 2a contains $N_L = 392 \times 392$ diffractive features, each with a size $225nm \times 225nm$, i.e., $\lambda_9/2\ \times \lambda_9/2$, which also corresponds to $\lambda_1/3.11\ \times \lambda_1/3.11$. Considering that our diffractive network operates based on traveling/propagating waves, the longer wavelengths experience reduced degrees of freedom due to the diffraction limit of light, which restricts the independent (useful) feature size on a diffractive layer to half of the wavelength in each spectral band.

Next, we further quantified the multispectral imaging quality provided by the diffractive network design shown in Fig. 2a using two additional performance metrics: Structural Similarity Index Measure (SSIM) and Peak Signal-to-Noise Ratio (PSNR); see the Methods section. Figure 2d illustrates the average SSIM and PSNR values achieved by the diffractive network as a function of the desired spectral bands. These image quality metrics were calculated between the diagonal images shown in Fig. 2b (the ground truth images on the left diagonal vs. the diffractive network output images on the right diagonal). Although there are some variations in the multispectral imaging quality of the diffractive network depending on the spectral band of the input light, the SSIM (PSNR) values have a very high lower bound (worst case performance) of 0.88 (19.8 dB). In addition, the mean SSIM and PSNR values are found as 0.93 and 22.06 dB, respectively. By summing up all the images in each column of Fig. 2b, we can create an image that visualizes the impact of the spectral cross-talk from the other $N_B - 1 = 8$ spectral channels on each target wavelength, which is shown at the bottom of the image matrix in Fig. 2b, as a separate row. Due to this spectral power cross-talk among channels (quantified in Fig. 2c), the average values of the SSIM and PSNR of the output multispectral image cube (computed across all the bands) drop to 0.65 and 16.24 dB, respectively. Also see Supplementary Fig. 1 for the cross-talk matrix and multispectral imaging performance of a diffractive multispectral imager designed for $N_B = 4$ spectral bands in the visible spectrum. Due to the reduced number of target spectral bands compared to the $N_B = 9$ case, the spectral power cross-talk is reduced for the $N_B = 4$ diffractive imager as quantified in Supplementary Fig. S1b; as a result, the diffractive network can synthesize multispectral image cubes with improved mean SSIM (0.82) and mean PSNR (19.29 dB) calculated across all the $N_B = 4$ bands.

To demonstrate diffractive multispectral imaging with an increased number of spectral channels, Figure 4a demonstrates the material thickness profiles of the diffractive layers constituting a new multispectral imager that was trained for $N_B = 16$, evenly distributed between $\lambda_{16} = 450$ nm to $\lambda_1 = 700$ nm, mapped onto a 4×4 monochrome pixel array repeating in space for snapshot multispectral imaging. Compared to the diffractive multispectral imager depicted in Fig. 2, this new diffractive design targets a lower spatial resolution due to the trade-off between $N_B$ and the spatial resolution of the snapshot multispectral imager. Similar to Fig. 2b, the output images on the diagonals of the multispectral image cube shown in Fig. 4b closely match the ground truth multispectral images at the input, highlighting the success of the diffractive imaging design. The off-diagonal images that are dark (see Fig. 4b) further illustrate the success of the spectral routing performed by the diffractive multispectral imager, minimizing the cross-talk among channels. Figure 4c also illustrates the average spectral signal contrast synthesized by the diffractive network at its output for $N_B = 16$ spectral bands. Compared to the signal contrast map of the previous diffractive network design ($N_B = 9$ shown in Fig. 2c), the values in Fig. 4c point to a slight decrease in the average spectral contrast at the output of this new diffractive multispectral imager with $N_B = 16$. However, the output image quality of the diffractive multispectral imager with $N_B = 16$ is still outstanding: the output SSIM (PSNR) values have a very good lower bound of 0.88 (19.62 dB), and the mean SSIM and PSNR values are 0.92 and 22.0 dB, respectively (see Fig. 4d). Same as in Fig. 2, these image quality metrics were calculated between the diagonal images shown in Fig. 4b (left vs. right). By summing up all the images in each column of Fig. 4b, we can create an image that visualizes the impact of the power cross-talk from the other $N_B - 1 = 15$ spectral bands on each target wavelength, which is shown at the bottom of the image matrix in Fig. 4b, as a separate row. As a manifestation of the spectral power cross-talk quantified in Fig. 4c, the average values of SSIM and PSNR of the output multispectral image cube drop to 0.60 and 15.33 dB, respectively, calculated across all the $N_B = 16$ target spectral channels. Furthermore, this diffractive multispectral imager with $N_B = 16$ can route the input spectral bands onto designated output pixels with an average power contrast that is 11.06× larger with respect to the mean power carried by the remaining $N_B - 1 = 15$ spectral channels. Figure 5b also reports the spectral responsivity curves of the 4x4 virtual filter array at the output image FOV of the diffractive network.

Next, to experimentally demonstrate the presented diffractive multispectral imaging framework, we designed a diffractive network that can process terahertz wavelengths. This terahertz-based diffractive multispectral imager uses $K = 3$ layers (see Fig. 6) to form a virtual filter array at its output plane with periodically repeating 2×2 spectral pixels targeting 0.375 THz, 0.4 THz, 0.425 THz and 0.45 THz ($i.e., N_B = 4$). For the input object 'U' shown in Fig. 6b, the demosaiced output images predicted by the numerical forward model of our diffractive terahertz multispectral imager are depicted in Fig. 7a. In the 4-by-4 image matrix shown in Fig. 7a, the diagonal images represent the correct match between the spectral content of the illumination and the corresponding demosaiced pixels within each 2×2 cell of the virtual filter array; in other words, they represent the channels of the multispectral image cube, while the off-diagonal images show the cross-talk between different spectral bands. To quantify the performance of our diffractive multispectral imager, we compared each spectral channel of the multispectral image cube predicted by the numerical forward model of our diffractive terahertz multispectral imager with respect to the ground-truth image of the input object 'U', which achieved PSNR values of 15.12 dB, 14.93 dB, 15.03 dB and 13.30 dB for the spectral bands at 0.375 THz, 0.4 THz, 0.425 THz and 0.45 THz, respectively. To compare our numerical results with their experimental counterparts, Figure 7b illustrates the experimentally measured multispectral imaging results obtained through the 3D-printed multispectral diffractive imager shown in Fig. 6c, which provided a decent agreement between our numerical and experimental multispectral images. Quantitative evaluation of the experimental multispectral imaging results reveals PSNR values of 13.02 dB, 13.71 dB, 13.02 dB and 12.64 dB PSNR at 0.375 THz, 0.4

THz, 0.425 THz and 0.45 THz, respectively. Compared to our numerical results, these PSNR values point to ~1–2 dB loss of image quality which can be largely attributed to the limited lateral resolution and potential misalignments of the 3D printed diffractive multispectral imager shown in Fig. 6c.

Beyond the multispectral image quality, we also quantified the spectral cross-talk performance of the experimentally tested diffractive multispectral imager. Figures 7c and 7d illustrate the spectral cross-talk matrices generated by the numerical forward model of the diffractive multispectral imager shown in Fig. 6b and its experimentally measured counterpart using the 3D-printed diffractive design shown in Fig. 6c, respectively. For a given virtual filter array designated to a particular spectral band, the ratio between the mean power of the target spectral band and the mean power of all the other 3 undesired spectral bands was found to be 2.42 (numerical) and 2.21 (experimental) based on the matrices shown in Fig. 7c and 7d, respectively, providing a decent agreement between the numerical and experimental (3D-fabricated) models of our diffractive multispectral imager.

In general, a key design parameter in diffractive optical networks is the number of diffractive features, $N$, that are engineered using deep learning since it directly determines the number of independent degrees of freedom in the system[55,56,62]. Figures 8a-c compare the multispectral imaging quality achieved by four different diffractive network architectures as a function of $N$ for $N_B = 4$, 9 and 16, respectively. For example, the diffractive multispectral imager designs for $N_B = 9$ and $N_B = 16$ shown in Figs. 2 and 4, respectively, contain in total $N = 392 \times 392 \times 5$ trainable diffractive features equally distributed over $K = 5$ diffractive layers, i.e., the number of diffractive features per layer is, $N_L = 392 \times 392$. While these two diffractive multispectral imagers can achieve average SSIM (PSNR) values of 0.93 (22.06 dB) and 0.92 (22.00 dB) at their output images, respectively, the diffractive multispectral imager architectures with fewer $N$ cannot match their performance. For instance, in the case of a diffractive multispectral imager design based on $N_B = 9$, $N_L = 196 \times 196$ and $K = 3$ (see Fig. 8b), the average output SSIM and PSNR values drop to 0.7 and 15.38 dB, respectively. Figure 8d further illustrates the impact of $N_B$ on the multispectral imaging performance of diffractive networks for four different combinations of $N_L$ and $K$. One can observe in Fig. 8d that for a fixed $N_L$ and $K$ combination, the multispectral imaging performance is inversely proportional to $N_B$, which is expected due to the increased level of spectral multiplexing. As a comparison, the average SSIM (PSNR) values attained by the diffractive multispectral imager with the smallest $N = 196 \times 196 \times 3$ increase from 0.7 (15.38 dB) to 0.78 (16.44 dB) when $N_B = 9$ is reduced to $N_B = 4$ spectral bands; this once again points to the relationship between $N$ and $N_B$, indicating that a larger $N_B$ would require additional diffractive degrees of freedom (a larger $N$) in order to perform the desired multispectral imaging task over a larger set of spectral bands.

Another critical figure of merit regarding the design of diffractive multispectral imagers is the power transmission efficiency of the optically synthesized virtual filter array. Figures 3c and 5C illustrate the power transmission efficiencies of the virtual filter arrays generated by the diffractive multispectral imager networks with $N_B = 9$ and $N_B = 16$ distinct bands within the visible spectrum. For example, based on the data depicted in Fig. 3c, the highest and lowest transmission efficiencies for $N_B = 9$, are found as 21.56% and 20.70% at 450 nm and 700 nm, respectively. On average, this diffractive multispectral imager can provide 20.96% virtual filter transmission efficiency for $N_B = 9$ spectral bands targeted by the 3×3 repeating cell of the virtual filter array. However, the deep learning-based training of this diffractive multispectral imager shown in Fig. 2 focused solely on the quality of the multispectral optical imaging, i.e., the output diffraction efficiency related training loss term ($\mathcal{L}_e$) was dropped (see the Supplementary Methods). While this training strategy drives the evolution of the diffractive surfaces to maximize the multispectral imaging performance, the associated virtual filter array transmission efficiency reflects only a lower performance bound that can be achieved by a diffractive multispectral imager with the same optical architecture. To find a better balance between the multispectral imaging quality and the power efficiency of the virtual filter array, the loss function that guides the diffractive multispectral imager design during its deep learning-based training can include an additional term, $\mathcal{L}_e$, penalizing poor diffraction efficiencies (see the Supplementary Methods section). The multiplicative constant, $\gamma$, determines the weight of the diffraction efficiency penalty, $\mathcal{L}_e$, controlling the trade-off between the multispectral imaging quality and the power efficiency of the associated virtual spectral filter array. To quantify the impact of $\mathcal{L}_e$ and $\gamma$ on the performance of diffractive multispectral imagers, we trained new diffractive models that share an identical optical architecture with the diffractive multispectral imager shown in Fig. 2 ($N_B = 9$ within the visible spectrum), where each design used a different value of $\gamma$. The results of this analysis are shown in Fig. 9, which indicate that it is possible to create a 5-layer diffractive multispectral imager with $N_B = 9$, achieving an average virtual filter transmission efficiency as high as 79.32%. Furthermore, the compromise in multispectral image quality in favor of this significantly increased power transmission efficiency turned out to be only minimal: while the average SSIM (PSNR) values achieved by the lower efficiency diffractive networks shown in Fig. 2 were 0.93 (22.06 dB), the more efficient diffractive multispectral imager design with 79.32% average virtual filter array transmission efficiency achieves an SSIM of ~0.91 and a PSNR of 21.42 dB (see Fig. 9).

Finally, we should emphasize that the presented diffractive multispectral imager design framework using deep learning-based optimization of phase-only diffractive layers can also be extended to spatially incoherent illumination. In the forward model of such a design to be trained using deep learning, each spatially incoherent wavefront at a given band can be decomposed into field amplitudes with random 2D input phase patterns, and the output image can be synthesized by averaging the intensities resulting from various independent random phase patterns for the same input field amplitude. The downside of such an incoherent multispectral imager design is that it would take much longer to converge using deep learning since each forward operation during the training phase would need many independent runs with random input phase patterns for each batch of the multispectral training input images. Therefore, at the cost of a longer one-time training effort, phase-only diffractive layers can also be optimized using deep learning to create a spatially incoherent snapshot multispectral imager, following the same design principles outlined in this work.

In summary, we demonstrated snapshot diffractive multispectral imagers that can create a virtual spectral filter array over the pixels of a monochrome focal-plane-array or image sensor without the need for a conventional filter array, while simultaneously establishing an imaging condition between the input and output fields-of-view. Owing to their extremely compact form factor, power-efficient optical forward operation (reaching >79% filter transmission efficiency) and high-quality spectral filtering capabilities, the presented diffractive multispectral imagers can be useful for numerous imaging and sensing applications, covering different parts of the spectrum where high-density and wide-area multispectral filter arrays are not readily available. Finally, since our diffractive designs are based on isotropic materials, their multispectral imaging capability and virtual spectral filter responses are independent of the input polarization state of the illumination light, which provides an additional, important advantage.

## 3. Methods

### 3.1 Training forward model of diffractive multispectral imagers

#### 3.1.1 Optical forward model

$D^2$NN framework uses deep learning to devise the transmission/reflection coefficients of diffractive features located over a series of optical modulation surfaces. The modulation coefficient over each diffractive feature/neuron is controlled through one or more physical design variables. The presented diffractive multispectral imagers in this work were designed to be fabricated based on a single dielectric material and the material thickness, $h$, was selected as the physical parameter for controlling the complex-valued modulation coefficient associated with each diffractive feature. For a given diffractive layer, the transmittance coefficient of a diffractive feature located on the $l^{th}$ layer at a coordinate of $(x_q, y_q, z_l)$ is defined as,

$$t(x_q, y_q, z_l) = \exp\left(\frac{-2\pi\kappa h(x_q, y_q, z_l)}{\lambda}\right)\exp\left(\frac{-j2\pi(n - n_m)h(x_q, y_q, z_l)}{\lambda}\right) \quad (1)$$

where $n$ and $\kappa$ denote the real and imaginary parts of the refractive index of the fabrication dielectric material, respectively, and $n_m = 1$ corresponds to the refractive index of the propagation medium (air) between the layers. In the case of the diffractive multispectral imagers designed to operate at the visible wavelengths, the material of the diffractive layers was selected as Schott glass of type 'BK7' due to its wide availability and low absorption[63]. Since its absorption coefficient for the visible spectrum is on the order of $10^{-3}$ cm$^{-1}$, the imaginary part of the refractive index was ignored, i.e., it was assumed to be absorption-free; considering the fact that our diffractive designs extend <45 µm in the axial direction, this is a valid assumption. For the experimentally tested diffractive multispectral imaging network shown in Figs. 6-7, on the other hand, the real and imaginary parts of the diffractive materials were measured experimentally using a THz spectroscopy system, i.e., $n = 1.6524, 1.6518, 1.6512, 1.6502$, and $\kappa = 0.05, 0.06, 0.06, 0.06,$ at 0.375, 0.400, 0.425 and 0.450 THz, respectively.

Each diffractive layer was modeled as a multiplicative thin modulation surface in the optical forward model. The light propagation between successive diffractive layers was implemented based on the Rayleigh-Sommerfeld scalar diffraction theory; since the smallest diffractive features considered here have a size of ~λ/2 this is a valid assumption for all-optical processing of diffraction-limited traveling/propagating fields, without any evanescent waves. According to this diffraction formulation, the free-space diffraction is interpreted as a linear, shift-invariant operator with an impulse response of,

$$w(x, y, z) = \frac{z}{r^2}\left(\frac{1}{2\pi r} + \frac{n}{j\lambda}\right)\exp\left(\frac{j2\pi nr}{\lambda}\right) \quad (2)$$

where $r = \sqrt{x^2 + y^2 + z^2}$. Based on Eq. 2, $q^{th}$ diffractive feature on the $l^{th}$ layer, at $(x_q, y_q, z_l)$, can be described as the source of a secondary wave, generating the field in the form of,

$$w_q^l(x, y, z) = \frac{z - z_l}{\left(r_q^l\right)^2}\left(\frac{1}{2\pi r_q^l} + \frac{n}{j\lambda}\right)\exp\left(\frac{j2\pi n r_q^l}{\lambda}\right) \quad (3)$$

where $r_q^l = \sqrt{\left(x - x_q\right)^2 + \left(y - y_q\right)^2 + (z - z_l)^2}$. These secondary waves created by the diffractive features on the diffractive layer $l$ propagate to the next layer, i.e., the $(l + 1)^{th}$ layer and are spatially superimposed. Accordingly, the light field incident on the $p^{th}$ diffractive feature at $(x_p, y_p, z_{l+1})$ can be written as $\sum_q A_q^l w_q^l(x_p, y_p, z_{l+1})$, where $A_q^l$ is the complex amplitude of the wave field right after the $q^{th}$ diffractive feature of the $l^{th}$ layer. This field is modulated through the field transmittance of the diffractive unit at $(x_p, y_p, z_{l+1})$, i.e., $t(x_p, y_p, z_{l+1})$, where a new secondary wave is generated, described by:

$$u_p^{l+1}(x, y, z) = w_p^{l+1}(x, y, z)\, t(x_p, y_p, z_{l+1}) \sum_q A_q^l w_q^l(x_p, y_p, z_{l+1}). \quad (4)$$

The outlined successive modulation and the secondary wave generation processes continue until the waves propagating through the diffractive network reach the output image plane. Although the forward optical model described by Eqs. 1-4 is given over a continuous 3D coordinate system, during our deep learning-based training of the presented diffractive optical networks, all the wave fields and the modulation surfaces were represented based on their discrete counterparts. For the diffractive multispectral imager designs operating at the visible bands, the spatial sampling rate was set to be $0.5\lambda_{N_B} = 225\ nm$ for both $N_B = 4\ and\ 9$, which was also equal to the size of a diffractive feature. For the experimentally tested diffractive multispectral imaging system, on the other hand, the sampling rate was selected as $0.375\lambda_{N_B}$ and the size of each diffractive feature was taken as $0.75\lambda_{N_B}$ with $N_B = 4$.

### 3.1.2 Design of the experimentally tested diffractive multispectral imager operating at terahertz bands

As shown in Fig. 6, the size of the input and output FOVs of the experimentally tested diffractive multispectral imager with $N_B = 4$ were set to be $37.5\lambda_1 \times 37.5\lambda_1$, where $\lambda_1$~0.8 mm is the wavelength at 0.375 THz. It was assumed that the THz output image plane has 100 ($10 \times 10$) pixels of size $3.75\lambda_1 \times 3.75\lambda_1$. Since $N_B = 4$, these $10 \times 10$ pixels were divided into groups of $2 \times 2$ virtual spectral filters repeating in space. The fabricated diffractive multispectral imager was trained using randomly generated intensity patterns, representing the amplitude transmission of the input objects. The 3D-printed blind test object is the letter 'U' designed based on a $5 \times 5$ binary image with each pixel corresponding to an area of $7.5\lambda_1 \times 7.5\lambda_1$.

The size of each diffractive feature on the 3D-printed diffractive layers shown in Fig. 6 equals ~0.5mm×0.5mm. Each of the 3 fabricated diffractive surfaces processes the incoming waves based on $100 \times 100$ optimized diffractive features, extending over $62.5\lambda_1 \times 62.5\lambda_1$. In the optical forward model of this diffractive network, all the axial distances between (1) the input FOV and the first diffractive surface, (2) two successive diffractive surfaces and (3) the last diffractive layer and the output FOV were set to be 40 mm, i.e., ~$50\lambda_1$. The variables $h_m$ and $h_b$ were taken to be $1.56\lambda_1$ and $0.625\lambda_1$, respectively (see the Supplementary Methods section).

The fabricated diffractive multispectral imager shown in Fig. 6 was trained based on $\mathcal{L}'$ with $\gamma = 0.15$ (see the Supplementary Methods). Based on this $\gamma$ value, the $K = 3$ layer diffractive optical network shown in Fig. 6 provides 5.68%, 5.32%, 5.2% and 5.01% virtual filter array transmission efficiency ($T$) for the spectral components at 0.375 THz, 0.4 THz, 0.425 THz and 0.45 THz, respectively.

The forward model of a 3D-printed diffractive network is prone to physical errors, e.g., layer-to-layer misalignments. To mitigate the impact of these experimental error sources, such misalignments were modeled as random variables and incorporated into the forward training model so that the deep learning-based evolution of the diffractive surfaces is enforced to converge to solutions that show resilience against implementation errors[53]. Accordingly, the diffractive network design shown in Fig. 6 was vaccinated against random 3D layer-to-layer misalignments in the form of lateral and axial translations as well as in-plane rotations. For this, we introduced 4 uniformly distributed random variables, ${D_x}^l$, ${D_y}^l$, ${D_z}^l$ and ${D_\theta}^l$, representing the random errors in the 3D location and orientation of a diffractive layer, $l$, i.e.,

$${D_x}^l \sim U(-\Delta_x, \Delta_x)$$

$${D_y}^l \sim U(-\Delta_y, \Delta_y)$$

$${D_z}^l \sim U(-\Delta_z, \Delta_z)$$

$${D_\theta}^l \sim U(-\Delta_\theta, \Delta_\theta)$$

where $\Delta_x$, $\Delta_y$, $\Delta_z$ and $\Delta_\theta$ denote the error range anticipated based on the fabrication margins of our experimental system. For the 3D-printed diffractive optical network shown in Fig. 6, the range of the random errors for the lateral misplacement of the diffractive surfaces was taken as $\Delta_x = \Delta_y = 0.625\lambda_1$. The variable, $\Delta_z$, which controls the maximum axial displacement of each layer, was set to be $2.5\lambda_1$. The range of errors in the orientation of each layer around the optical axis was assumed to be within $(-2°, 2°)$, i.e., $\Delta_\theta = 2°$. During the training stage, ${D_x}^l$, ${D_y}^l$, ${D_z}^l$ and ${D_\theta}^l$ were updated for each layer, $l$, independently for every batch of input objects, introducing a new set of random misalignment errors into the forward optical model at each error backpropagation step.

The numerically computed and experimentally measured power cross-talk matrices shown in Figs. 7c-d, were computed based on the images of the letter 'U' at 4 different illumination wavelengths: ~0.8 mm, ~0.75 mm, ~0.7 mm and ~0.66 mm.

### 3.1.3 Details of the experimental setup

The schematic diagram of the experimental setup is given in Fig. 6. In this system, the THz wave incident on the object was generated through a horn antenna compatible with the source WR2.2 modular amplifier/multiplier chain (AMC) from Virginia Diode Inc. (VDI). Electrically modulated with a 1 kHz square wave, the AMC received an RF input signal that is a 16 dBm sinusoidal waveform at 11.111 GHz ($f_{RF1}$). This RF signal is multiplied 34, 36, 38 and 40 times to generate a continuous-wave (CW) radiation at ~0.375 THz, ~0.4 THz, ~0.425 THz and ~0.45 THz, corresponding to ~0.8 mm, ~0.75 mm, ~0.7 mm and ~0.66 mm in wavelength, respectively. The exit aperture of the horn antenna was placed ~60 cm away from the object plane of the 3D-printed diffractive optical network so that the beam profile of the THz illumination closely approximates a uniform plane wave. The diffracted THz light at the output plane was collected using a single-pixel Mixer/AMC from Virginia Diode Inc. (VDI). A 10 dBm sinusoidal signal at 11.083 GHz was sent to the detector as a local oscillator for mixing so that the down-converted signal is at 1GHz. The $37.5\lambda_1 \times 37.5\lambda_1$ output FOV was scanned by placing the single-pixel detector on an XY stage that was built by combining two linear motorized stages (Thorlabs NRT100). The scanning step size was set to be 1 mm~$1.25\lambda_1$. The down-converted signal of a single-pixel detector at each scan location was sent to low-noise amplifiers (Mini-Circuits ZRL-1150-LN+) to amplify the signal by 80 dBm and a 1 GHz (+/-10 MHz) bandpass filter (KL Electronics 3C40-1000/T10-O/O) to clean the noise coming from unwanted frequency bands. Following the amplification, the signal was passed through a tunable attenuator (HP 8495B) and a low-noise power detector (Mini-Circuits ZX47-60), and then the output voltage was read by a lock-in amplifier (Stanford Research SR830). The modulation signal was used as the reference signal for the lock-in amplifier and accordingly, we conducted a calibration by tuning the attenuation and recording the lock-in amplifier readings. The lock-in amplifier readings at each scan location were converted to a linear scale according to the calibration.

The diffractive multispectral imager was fabricated using a 3D printer (Objet30 Pro, Stratasys Ltd.). The optical architecture of the 3D-printed diffractive optical network consisted of an input object and 3 diffractive layers (see Fig. 6). While the active modulation area of our 3D-printed diffractive layers was 5 cm × 5 cm ($62.5\lambda_1 \times 62.5\lambda_1$), they were printed as light-modulating insets surrounded by a uniform slab of the printing material with a thickness of 2.5 mm.

### 3.1.4 Training details and image quality metrics

The image quality metrics SSIM and PSNR were computed based on the comparison between the low-resolution ground-truth image cube, $I_{GT,LR}[k,r,w]$, and the output image cube formed through the demosaicing of the optical intensity patterns collected by the image sensor, $I_{S,LR}[k,r,w]$. Both PSNR and SSIM metrics were computed separately for each spectral channel. The PSNR achieved by a diffractive multispectral imager for the spatial information in a spectral band, $w'$, was computed based on,

$$PSNR_{w'} = 20log_{10}\left(\frac{1}{\sqrt{\sum_k \sum_r \left|I_{GT,LR}[k,r,w'] - I_{S,LR}[k,r,w']\right|^2}}\right),$$

To compute the SSIM metric, we used the built-in tf.image.ssim() function in TensorFlow based on its default parameters. Each data point in SSIM and PSNR values shown in Figs. 2 and 4 represents the average value calculated using 2080 blind test objects created in a way that the amplitude channel of the spatial transmission function at each spectral band was modeled based on an image randomly selected from the 18.8K test images of the EMNIST dataset.

The deep learning-based training of the diffractive networks was implemented using Python (v3.6.5) and TensorFlow (v1.15.0, Google Inc.). The backpropagation updates were calculated using the Adam optimizer[64], and its parameters were taken as the default values in TensorFlow and kept identical in each model. The learning rates of the diffractive optical networks were set to be 0.001. The training batch size was taken as 8 during the deep learning-based training of all the presented diffractive multispectral imagers. The training of a 5-layer diffractive multispectral imager network with $392 \times 392$ diffractive features per layer (for 100 epochs) takes approximately 2 weeks using a computer with a GeForce GTX 1080 Ti Graphical Processing Unit (GPU, Nvidia Inc.) and Intel® Core ™ i7-8700 Central Processing Unit (CPU, Intel Inc.) with 64 GB of RAM, running Windows 10 operating system (Microsoft). Although the training time for the deep learning-based design of a diffractive multispectral imager is relatively long, it should be noted that this is a one-time effort. Once the diffractive multispectral imager is fabricated following the training stage, its physical forward optical operation consumes no power except the illumination beam.

**Supplementary Materials**

This file contains:

- Supplementary Figure S1.
- Design of diffractive multispectral imagers

# Figures

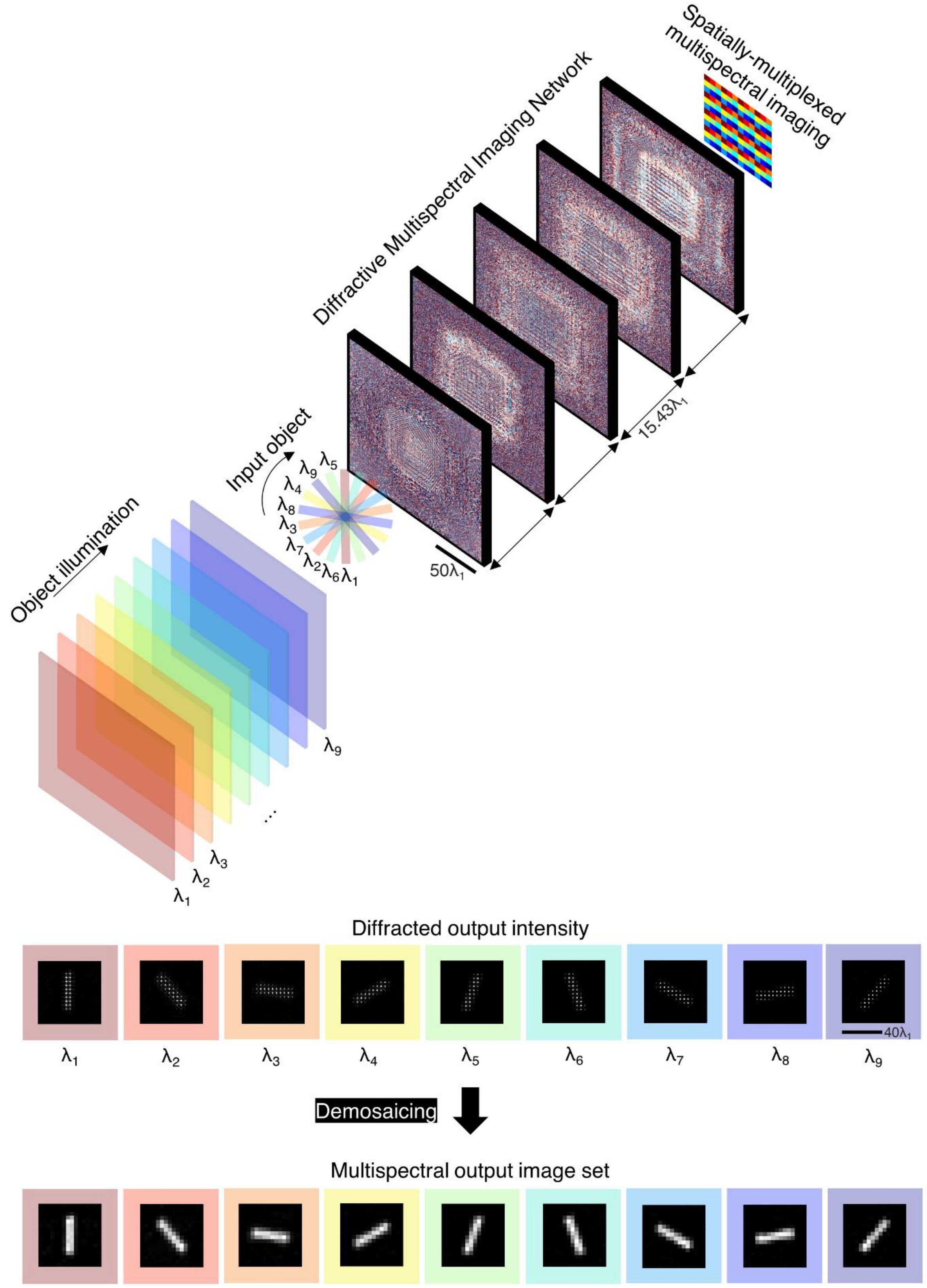

**Fig. 1: The schematic of a diffractive multispectral imager.** The depicted diffractive optical network simultaneously performs coherent optical imaging and spectral routing/filtering to achieve multispectral imaging by creating a periodic virtual filter array at the output. In this example, 3x3=9 spectral bands per virtual filter array are illustrated; Figures 4-5 report 4x4=16 spectral bands per virtual filter array. In alternative implementations, the diffractive multispectral network can also be placed behind the image plane of a camera (before the image sensor), transferring the multispectral image of an object onto the plane of a monochrome image sensor.

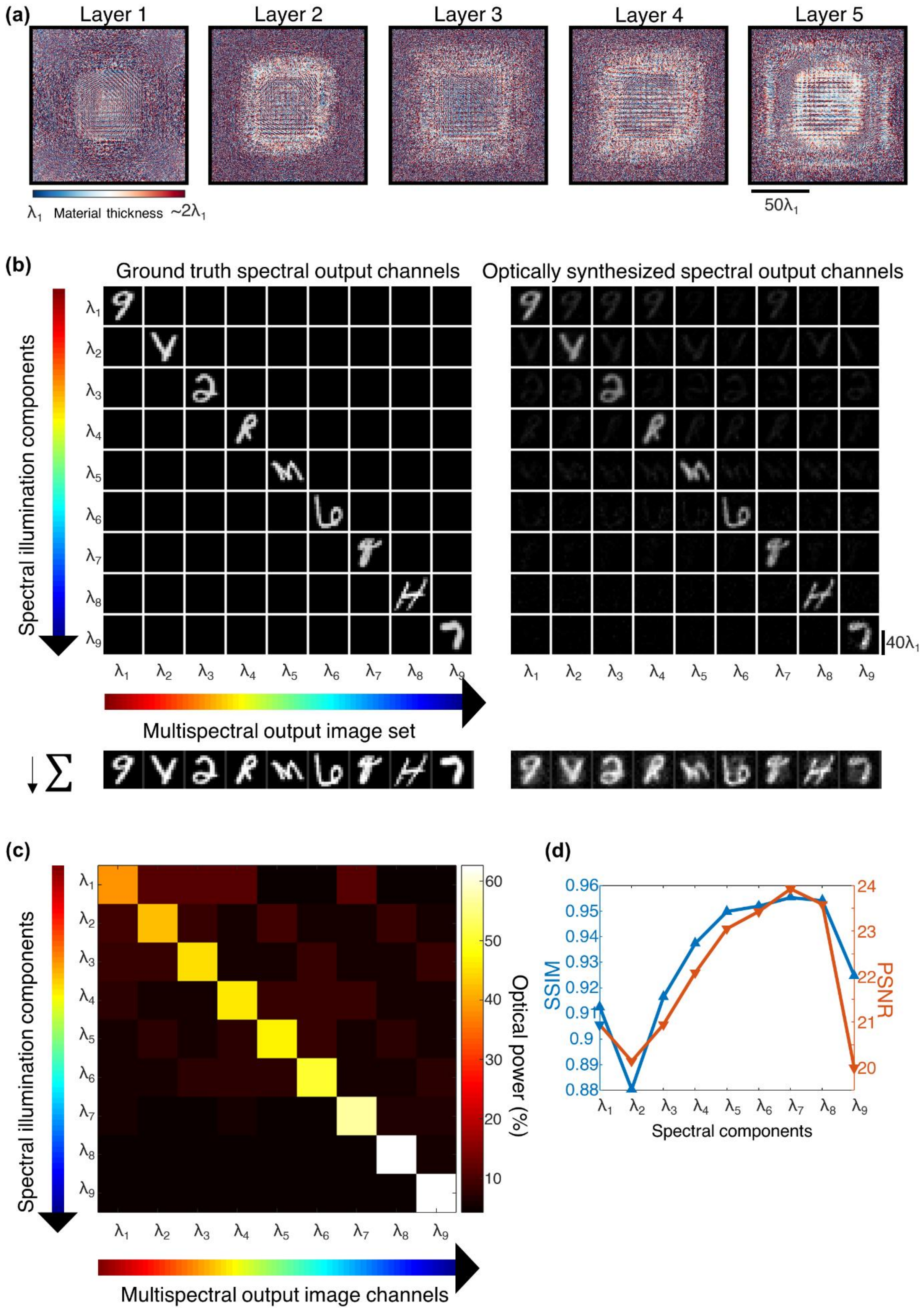

**Fig. 2. Performance of a diffractive multispectral imager with $N_B = 9$ spectral bands. a** The material thickness distribution of the diffractive layers trained using deep learning to spatially separate 9 distinct spectral bands, creating a periodic virtual filter array. **b** Cross-talk image matrix showing the output images at different illumination wavelengths. Off-diagonal images indicate that the level of spectral cross-talk is minimal. By summing up all the images in each column, the impact of the spectral cross-talk from the other 8 spectral channels on each target wavelength is visualized at the bottom of the image matrix, as a separate row. **c** Output optical power distribution as a function of the illumination wavelength. Each row in this matrix adds up to 100%, and the off-diagonal optical power percentages indicate the level of spectral cross-talk between different bands. **d** SSIM and PSNR values of the resulting images at the output of the diffractive network; these image quality metrics were calculated between the diagonal images shown in Fig. 2b (the ground truth images on the left diagonal vs. the diffractive network output images on the right diagonal).

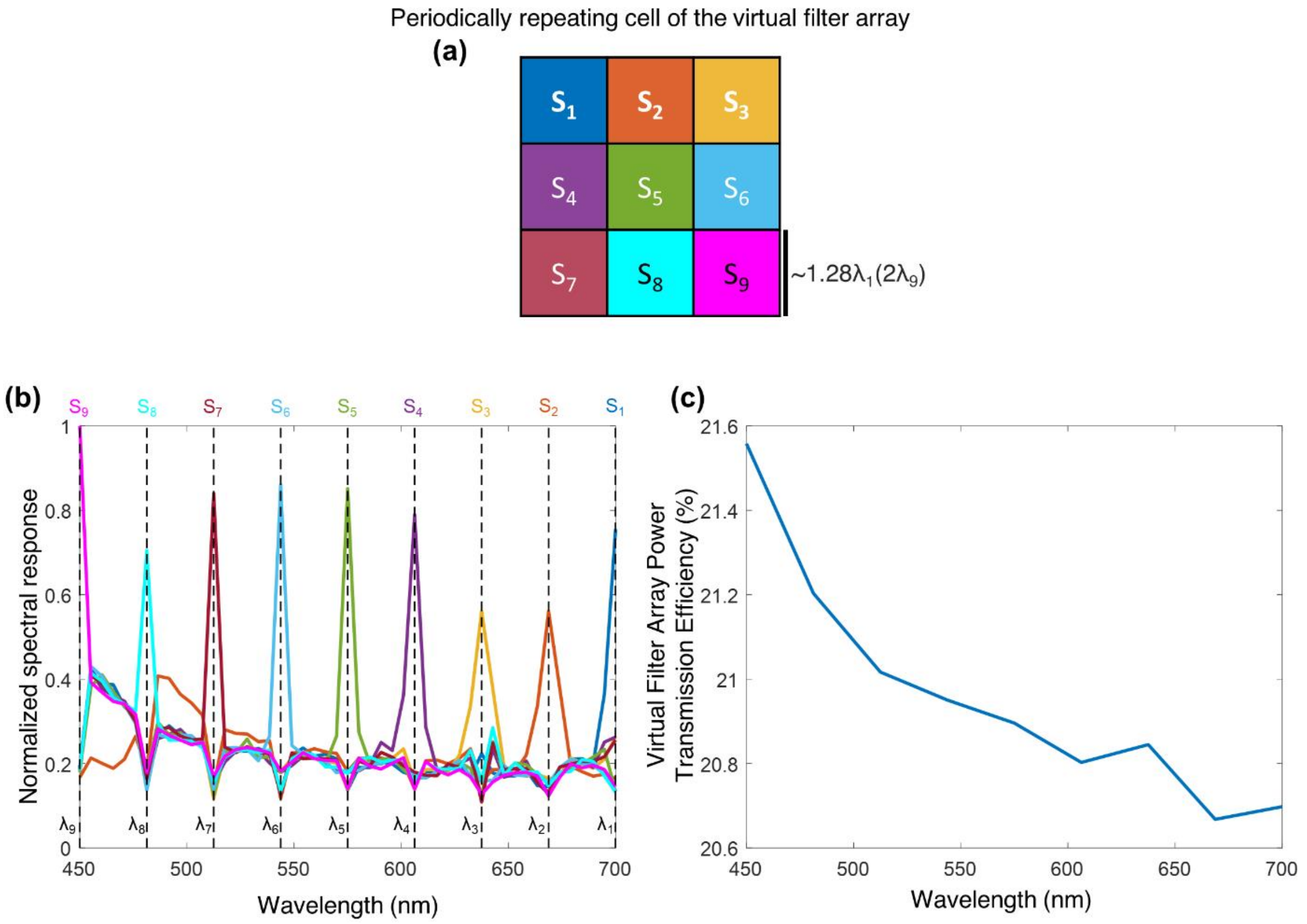

**Fig. 3. Spectral responsivity and power efficiency achieved by the diffractive multispectral imager shown in Fig. 2 ($N_B = 9$). a** The virtual spectral filter array periodically repeats at the output field of view of the diffractive network. The colored labels, $S_i$, $i = 1,2,3, \ldots ,9$, denote the virtual pixels assigned to the wavelength $\lambda_i$. **b** Average normalized spectral response of the virtual filter array. **c** The wavelength-dependent transmission power efficiency of the virtual filter array created by the diffractive optical network.

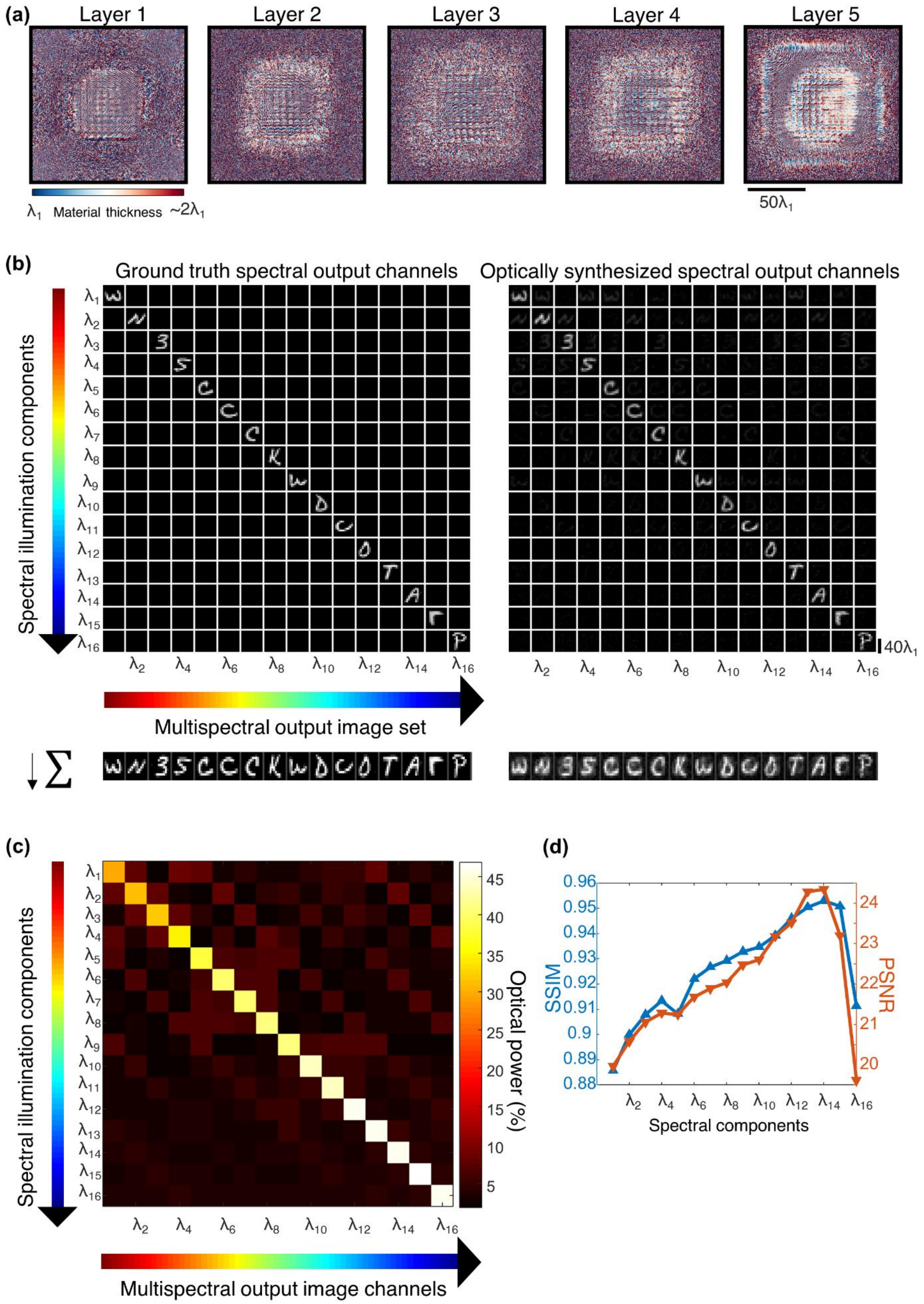

**Fig. 4. Performance of a diffractive multispectral imager with $N_B = 16$ spectral bands. a** The material thickness distribution of the diffractive layers trained using deep learning to spatially separate 16 distinct spectral bands, creating a periodic virtual filter array. **b** Cross-talk image matrix showing the output images at different illumination wavelengths. Off-diagonal images indicate that the level of spectral cross-talk is minimal. By summing up all the images in each column, the impact of the spectral cross-talk from the other 15 spectral channels on each target wavelength is visualized at the bottom of the image matrix, as a separate row. **c** Output optical power distribution as a function of the illumination wavelength. Each row in this matrix adds up to 100%, and the off-diagonal optical power percentages indicate the level of spectral cross-talk between different bands. **d** SSIM and PSNR values of the resulting images at the output of the diffractive network; these image quality metrics were calculated between the diagonal images shown in Fig. 4b (the ground truth images on the left diagonal vs. the diffractive network output images on the right diagonal).

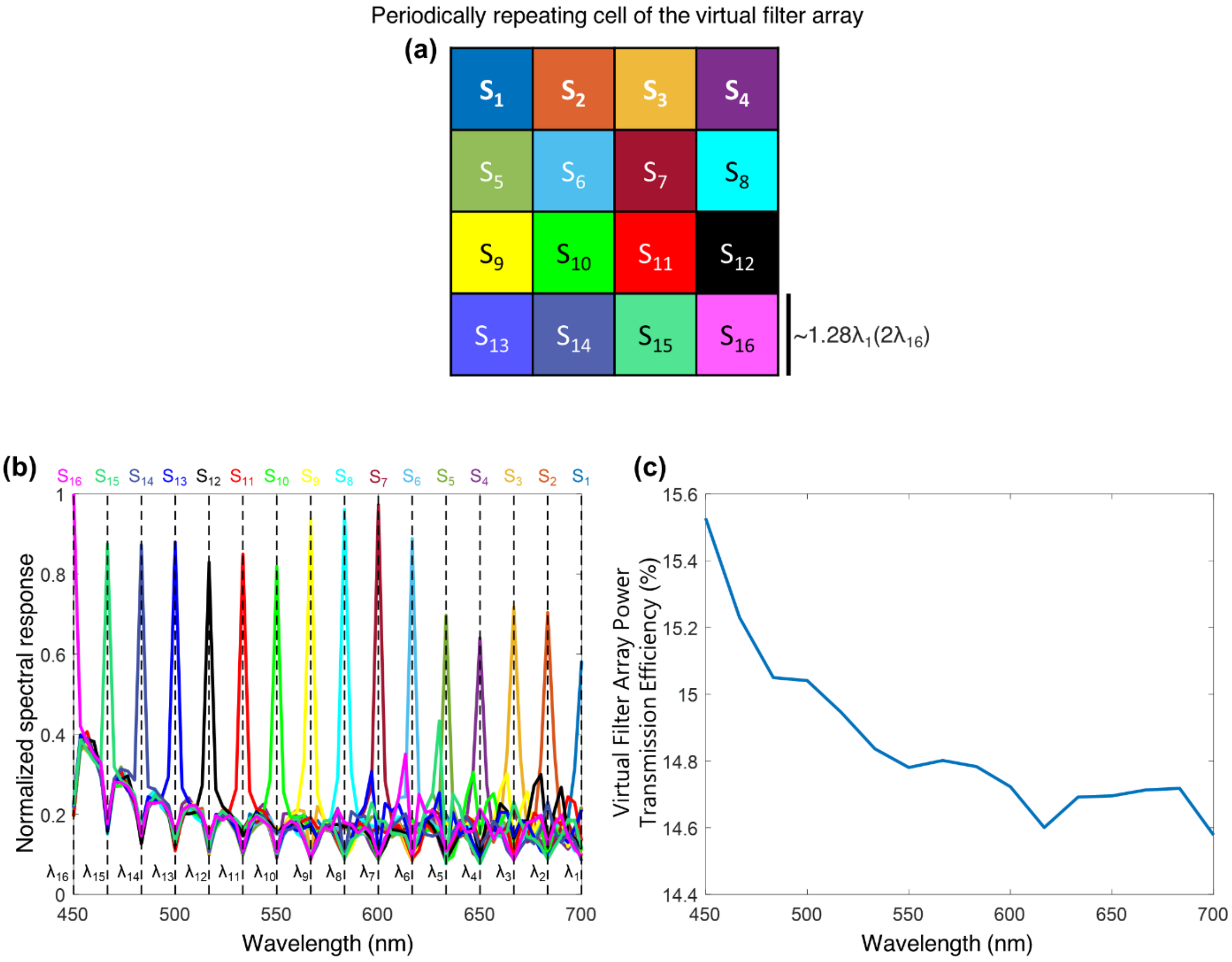

**Fig. 5. Spectral responsivity and power efficiency achieved by the diffractive multispectral imager shown in Fig. 4 ($N_B = 16$). a** The virtual spectral filter array periodically repeats at the output field of view of the diffractive network. The colored labels, $S_i$, $i = 1,2,3,\dots,16$, denote the virtual pixels assigned to the wavelength $\lambda_i$. **b** Average normalized spectral response of the virtual filter array. **c** The wavelength-dependent transmission power efficiency of the virtual filter array created by the diffractive optical network.

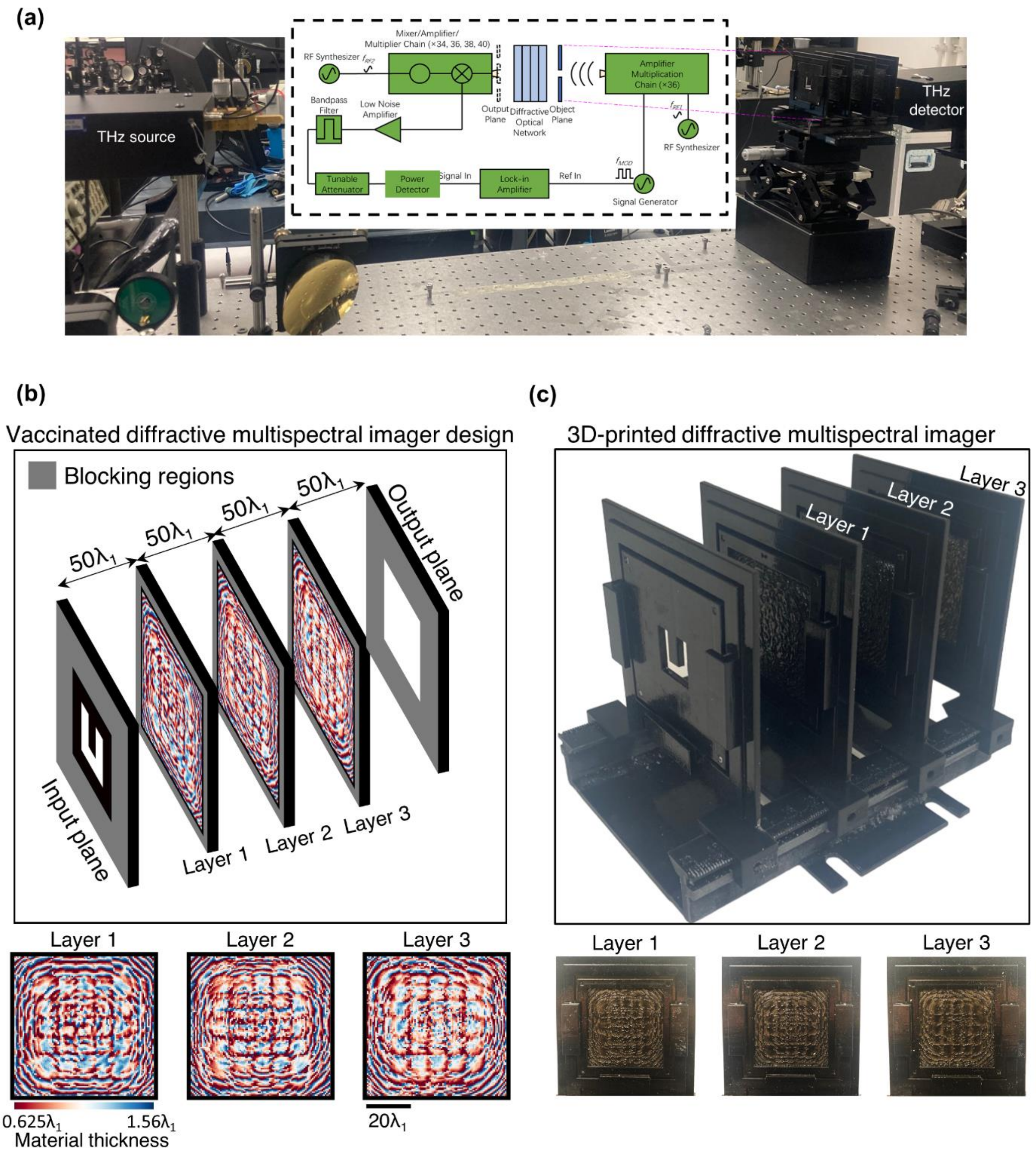

**Fig. 6. Experimental setup of the multispectral terahertz imager ($N_B = 4$). a** Schematic of the experimental setup using terahertz illumination and signal detection. **b** Optical design layout of the fabricated diffractive multispectral imager and the material thickness profiles of the 3 diffractive surfaces. **c** Fabricated diffractive network and the input object, the letter 'U'.

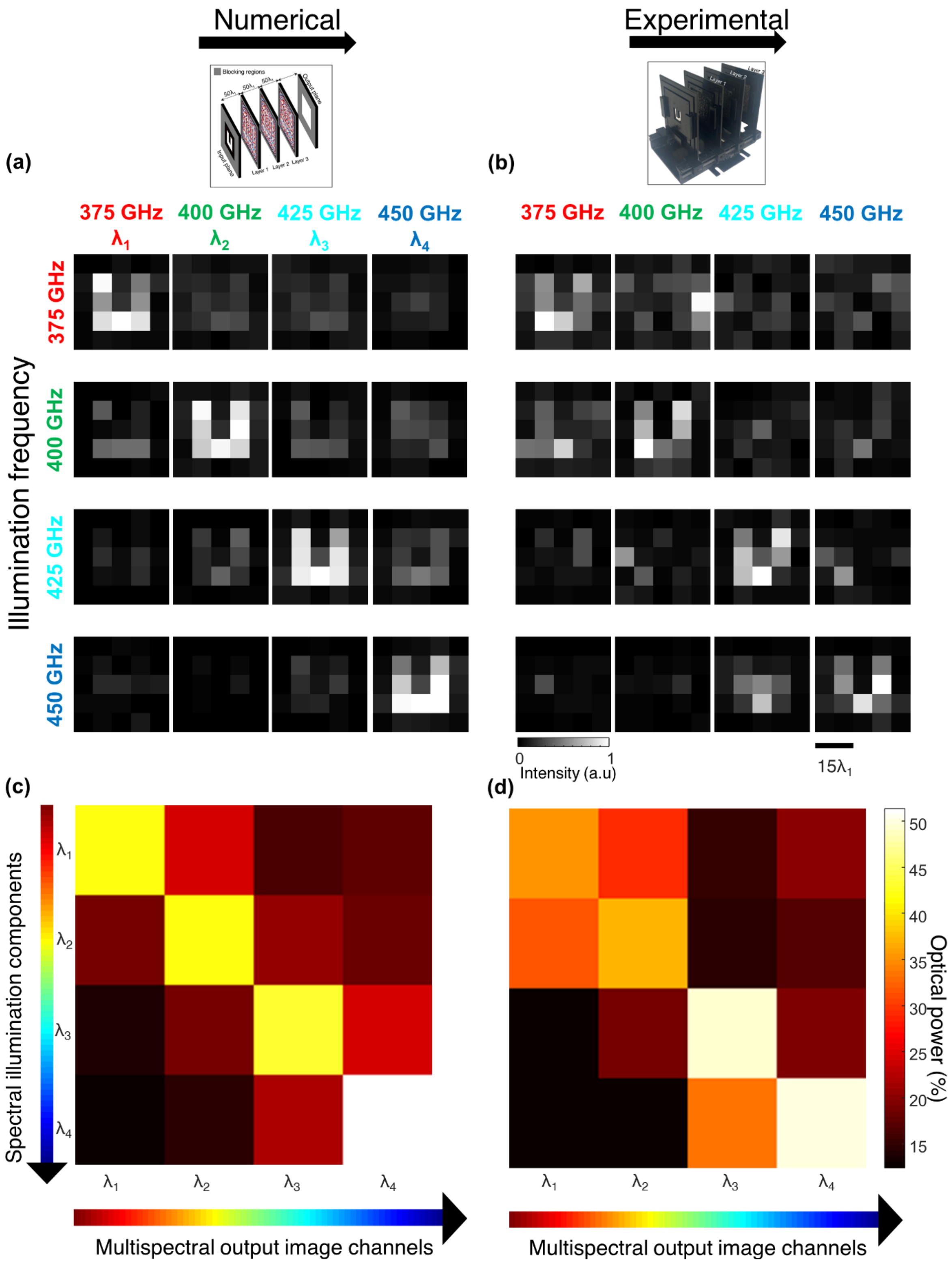

**Fig. 7. Experimental results for the diffractive multispectral imager shown in Fig. 6.** Multispectral imaging of letter 'U' at 4 distinct wavelengths in terahertz part of the spectrum based on the 3-layer diffractive optical network shown in Fig. 6. **a** Multispectral image cube synthesized by the numerical forward model of the diffractive optical network, after the demosaicing step. **b** Same as **(a)**, except that the images are extracted from the experimentally measured output optical intensity profiles. **c** Cross-talk matrix predicted by the numerical forward model of the diffractive multispectral imager in response to the input object 'U'. **d** Same as **(c)**, except that the entries in the cross-talk matrix represent the experimentally measured percentages of the optical power for each spectral band.

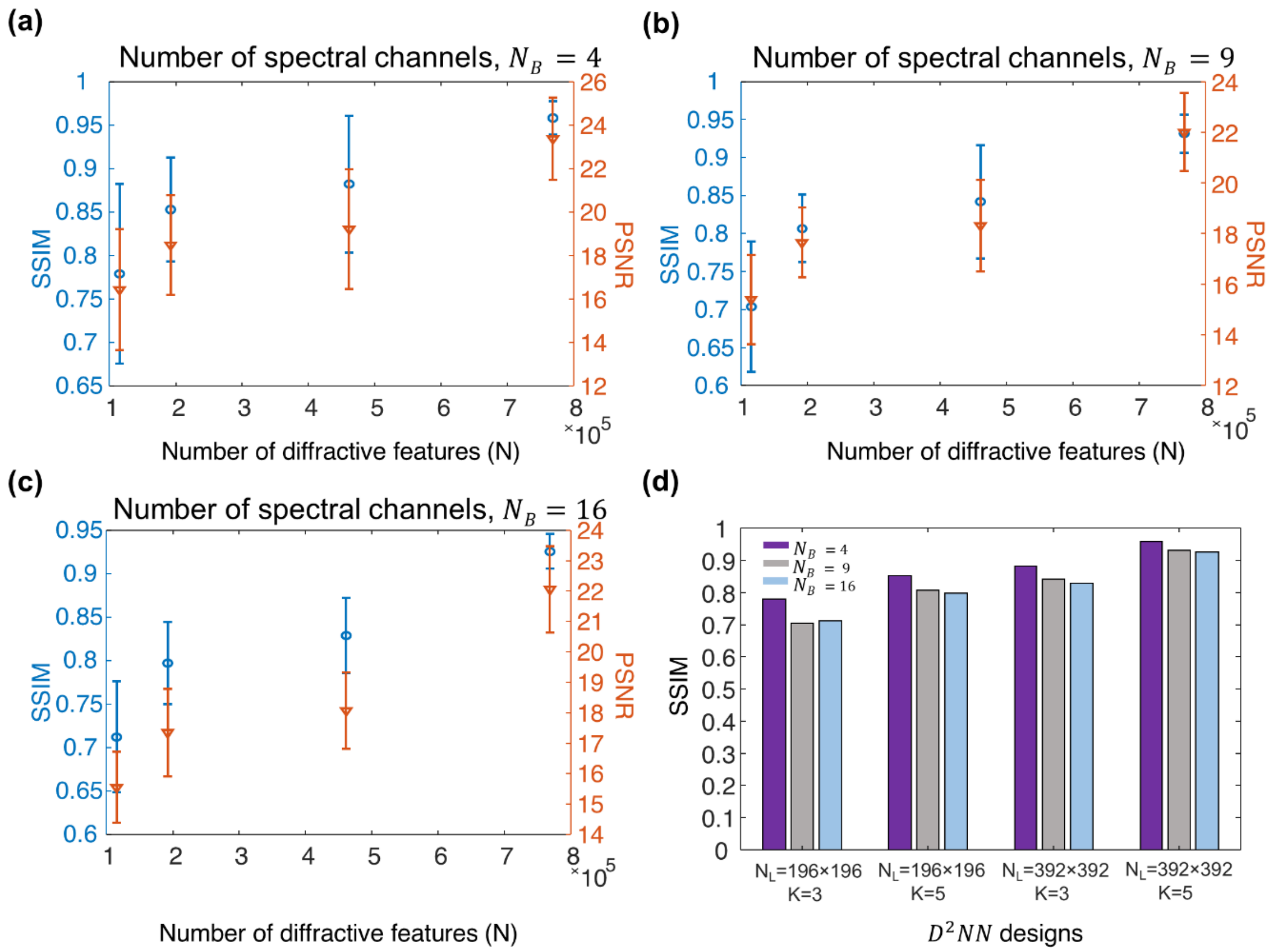

**Fig. 8. The impact of the number of diffractive features ($N$) and the targeted spectral bands ($N_B$) on the quality of the multispectral imaging using a diffractive optical network.** Mean SSIM and PNSR values as a function of $N$ are reported for **a** $N_B = 4$, **b** $N_B = 9$, **c** $N_B = 16$ spectral channels, forming a spatially repeating virtual spectral filter array. **d** The impact of $N_B$ on the SSIM of the output multispectral images for different diffractive network architectures. $K$ refers to the number of successive diffractive layers jointly trained for multispectral imaging, and $N_L$ refers to the number of trainable diffractive features per diffractive layer.

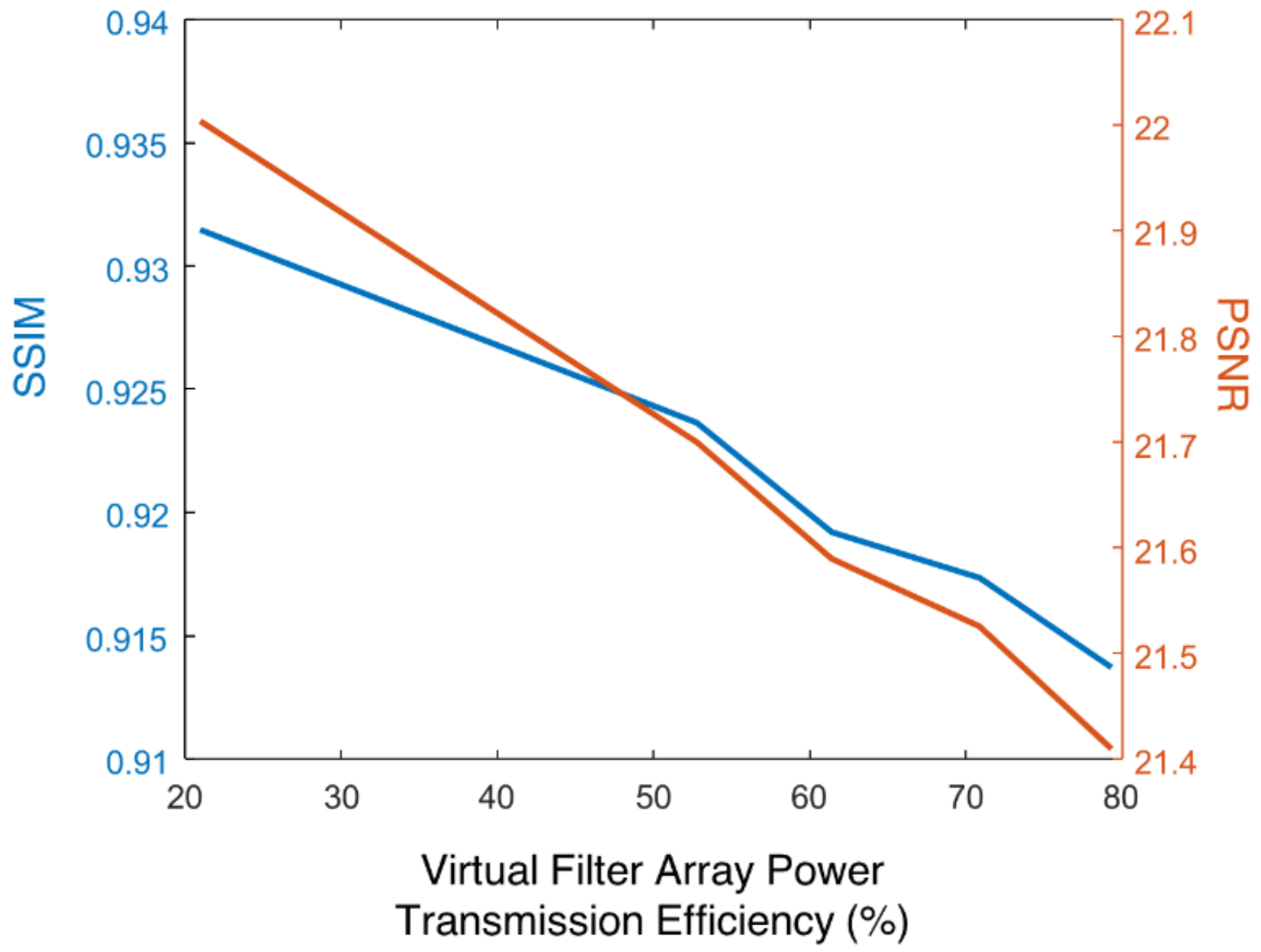

**Fig. 9. The trade-off between the multispectral imaging quality and the virtual filter array power transmission efficiency.** The impact of the additional power efficiency-related penalty term, $\mathcal{L}_e$ and its weight $\gamma$, on the multispectral image cube synthesized by the diffractive multispectral imagers that were trained to form 3×3 virtual spectral filter arrays assigned to $N_B = 9$ unique spectral bands within the visible spectrum. The optical architectures of these diffractive multispectral imagers are identical to the layout depicted in Fig. 1. We report an average virtual filter transmission efficiency of >79% across $N_B = 9$ spectral bands with a minimal penalty on the output image quality.